\documentclass[natbib209]{emulateapj}
 \usepackage{psfrag}
 \shorttitle{Mass Loss in LMC Cepheids}
 \shortauthors{Neilson et al.}
  
\begin{document}

 \title{Testing Mass Loss in Large Magellanic Cloud Cepheids using Infrared and Optical Observations}
 \author{Hilding R. Neilson}
\affil{Department of Astronomy \& Astrophysics, University of Toronto, 50 St. George Street, Toronto, ON, Canada M5S 3H4}
\email{neilson@astro.utoronto.ca}
 \author{Chow--Choong Ngeow}
\affil{Department of Astronomy, University of Illinois, Urbana, IL 61801}
 \author{Shashi M. Kanbur}
\affil{State University of New York at Oswego, Oswego, NY 13126}
\author{John B. Lester}
\affil{University of Toronto Mississauga, Mississauga, ON, Canada L5L 1C6}

\begin{abstract}
It has been claimed that Period--Luminosity relations derived from infrared observations of Large Magellanic Cloud (LMC) Cepheids are less dependent on the metallicity of the Cepheids.  In this work, infrared observations of LMC Cepheids from the SAGE survey are combined with OGLE II optical observation to model and predict mass--loss rates.  The mass--loss rates are fit to the data and are predicted to range from about $10^{-12}$ to $10^{-7}M_\odot/yr$; however, the rates depend on the assumed value of the dust--to--gas ratio. By comparing the relations derived from observations to the relations derived from predicted infrared stellar luminosities from the mass--loss model, it is shown mass loss affects the structure and scatter of the infrared Period--Luminosity relation.  Mass loss produces shallower slopes of the infrared relations and a lower zero point.  There is also evidence for non--linearity in the predicted Period--Luminosity relations, and it is argued that mass loss produces larger infrared excess at lower periods, which affects the slope and zero point, making the PL relations more linear in the wavelength range of $3.6$ to $5.8$ $\mu m$.  Because the dust--to--gas ratio is metallicity dependent and mass loss may have a metallicity dependence, infrared Period--Luminosity relations have additional uncertainty due to metallicity.
\end{abstract}

\section{Introduction}
Cepheids are powerful standard candles because they follow a Period--Luminosity (PL) relation. This relation has been determined using Cepheids in the Large Magallanic Cloud (LMC) in optical and near infrared bands \citep{Laney1994}.  LMC Cepheids also provide insight into stellar astrophysics as they have lower metallicity relative to Galactic Cepheids, which has an effect on the pulsation of Cepheids. 

 Recently, infrared PL relations have been derived using Spitzer observations from the SAGE program \citep{Meixner2006} by \cite{Ngeow2008} and by \cite{Freedman2008}.  These infrared PL relations are important for extragalactic studies, and this will be even more so when the James Web Space Telescope begins operation. The infrared PL relations are powerful tools because metallicity does not contribute significantly \citep{Freedman2008} and because the pulsation amplitude decreases in the infrared.  However, IRAS observations have found infrared excesses in Galactic Cepheids \citep{Deasy1988}. Interferometric observations have also detected the existence of circumstellar envelopes around a number of Galactic Cepheids \citep{Kervella2006, Merand2006, Merand2007} in the K--band.   The observations of infrared excess imply that there may be an additional uncertainty in infrared PL relations,  and the excesses may play a role in LMC Cepheids.

One proposed mechanism for generating circumstellar shells and infrared excesses about Galactic Cepheids is a stellar wind similar to those generated from other evolved pulsating stars.  Mass loss is believed to generate shells about asymptotic giant branch stars [where the mass--loss rates are related to pulsation and dust condensation in the atmospheres, in both the Milky Way and the LMC \citep{Mattsson2008}].  One would not expect dust to form in the atmospheres of Cepheids because the temperatures are greater than $1500$ $K$, and dust--driving is not a plausible mechanism to generate a stellar wind. However, it has been argued that Galactic Cepheids use pulsation and shocks generated by pulsation to eject mass \citep{Willson1989}.  \cite{Neilson2008} developed an analytic model to study the affect of pulsation and shocks, in combination with radiative--line driving, on mass--loss rates in Galactic Cepheids, predicting rates of the order $10^{-10}$ to $10^{-7} M_\odot /yr$.  It is argued that at some large distance from the surface of the Cepheid, the wind cools enough that a small fraction of the gas condenses into a dust shell that produces an infrared excess. The predicted infrared excesses from the theoretical model are consistent with both interferometric observations and IRAS observations \citep{Deasy1988}.  The analytic model was also applied to theoretical models of Cepheids with the metallicities of the Small and Large Magellanic Clouds, and the Galaxy, and mass loss was found to be significant in Magellanic Cloud Cepheids as well \citep{Neilson2008b}.  This result, however, has not been tested by observations.

The dust shells that surround Cepheids are optically thin because they form at large distances from the Cepheids.  For instance, a Cepheid with $T_{\rm{eff}} = 6000$ $K$ will have a condensation radius of $r_c/ R_* = 0.5(T_c/T_{\rm{eff}})^{-5/2} = 16$, assuming a condensation temperature of $1500$ $K$.  At the distance where dust forms, the dust shells are optically thin, and do not contribute to the extinction of starlight.

Mass loss may also be a solution to the Cepheid mass discrepancy problem. The Cepheid mass discrepancy is the difference between Cepheid mass estimates based on stellar evolution isochrones and those based on stellar pulsation calculations.  The mass discrepancy is about $10$--$20\%$ in the Milky Way \citep{Caputo2005}, about $17$--$25\%$ in the LMC and about $20\%$ in the SMC \citep{Brocato2004, Keller2006}.  \cite{Keller2006} determined that the mass discrepancy increases as the metallicity decreases, and \cite{Caputo2005} argue that the discrepancy may also be a function of mass, though \cite{Keller2008} provides evidence against this result.  For mass loss to be a solution to the mass discrepancy, a Cepheid that starts on its first crossing with mass of $5M_\odot$ would need to lose about $1M_\odot$ or have an average mass--loss rate in the range of $10^{-8}$ to $10^{-6}M_\odot /yr$.

The purpose of this work is to model infrared excess in LMC Cepheids using SAGE observations in the IRAC bands combined with OGLE II observations of B,V, and I \citep{Udalski1999a, Udalski1999b}.  The next section outlines the observations and the model describing circumstellar dust created in a stellar wind that causes infrared excess.  The process for determining the mass--loss rates is also described.  The results are given in Section 3 and the predicted infrared PL relations are described in Section 4. The fifth section will explore possible driving mechanisms for mass loss in LMC Cepheids, testing if the mass loss behavior is similar to that proposed in \cite{Neilson2008, Neilson2008b}.

\section{The Data and Mass Loss Model}
\begin{figure}[t]
	\begin{center}
	\epsscale{1.25}
		\plotone{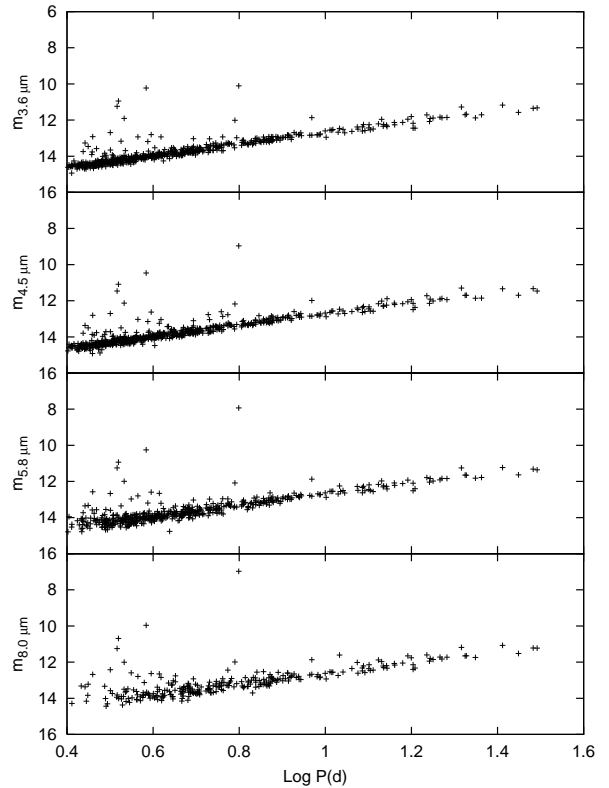}
\caption{The observed brightnesses of LMC Cepheids in the infrared as a function of period.}
	\label{f1}
	\end{center}
\end{figure}

We use OGLE II and SAGE observations of LMC Cepheids to determine mass--loss rates.  The OGLE II data are for B, V and I magnitudes while the SAGE magnitudes are in IRAC bands at wavelengths 3.6, 4.5, 5.8 and 8.0 $\mu m$. The SAGE data we used are adopted from \cite{Ngeow2008}, which consist of      
~730 OGLE II LMC Cepheids with $\log P > 0.4$. However we only use 488 of these Cepheids        
that have at least 3 IRAC bands and 2 of the BVI bands from this dataset.  The SAGE data are compiled by matching the position of OGLE II Cepheids with positions of infrared sources in the SAGE observations.  \cite{Ngeow2008} match sources if they are within $3.5$ arcseconds of the position of the OGLE II Cepheids.  Figure \ref{f1} show the infrared PL relations constructed using the IRAC magnitudes.  A number of (mostly short period) Cepheids appear to deviate from the infrared PL relations, implying there is some infrared excess.  

\begin{figure*}[t]
	\begin{center}
	\epsscale{1.}
		\plottwo{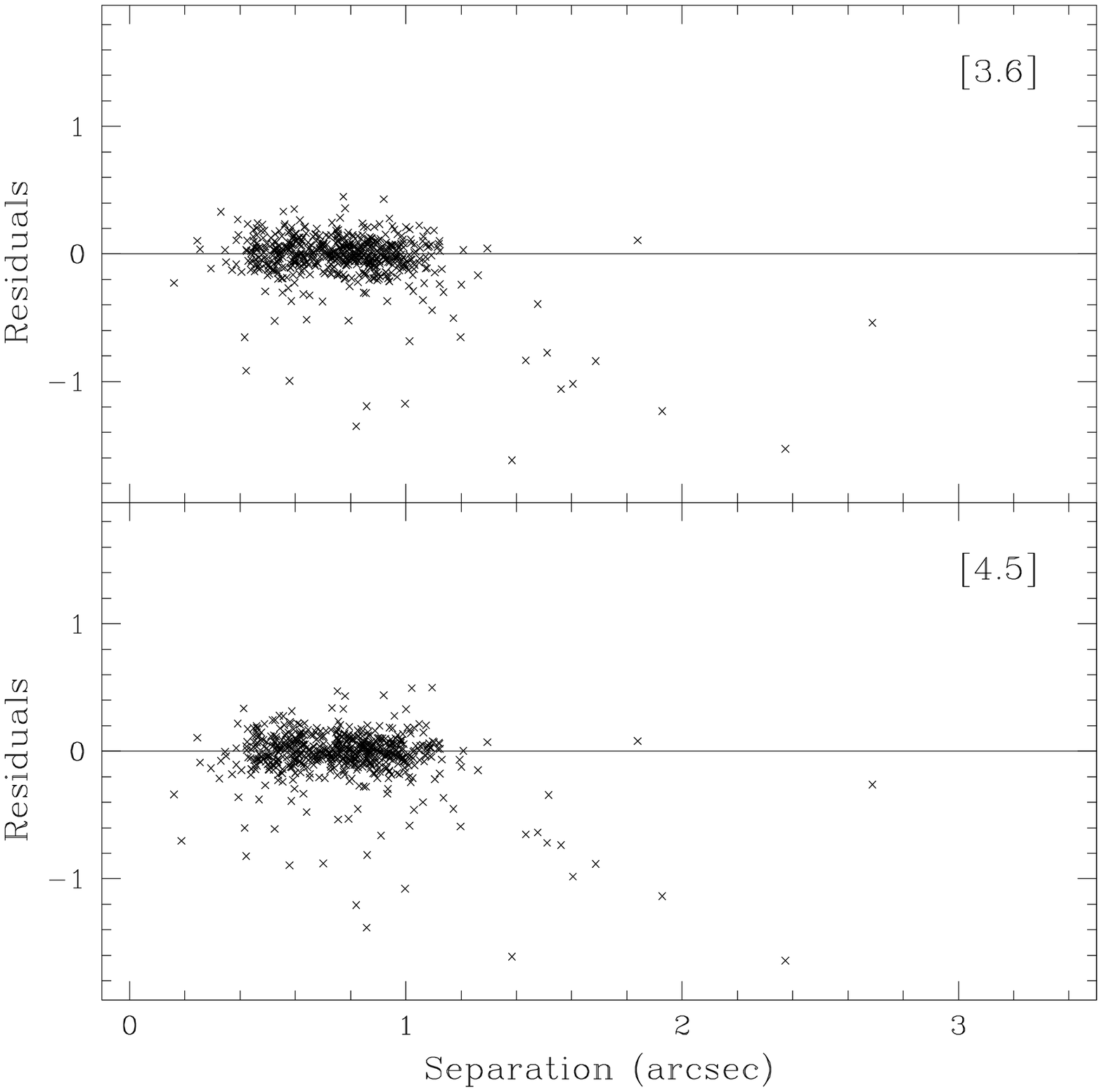}{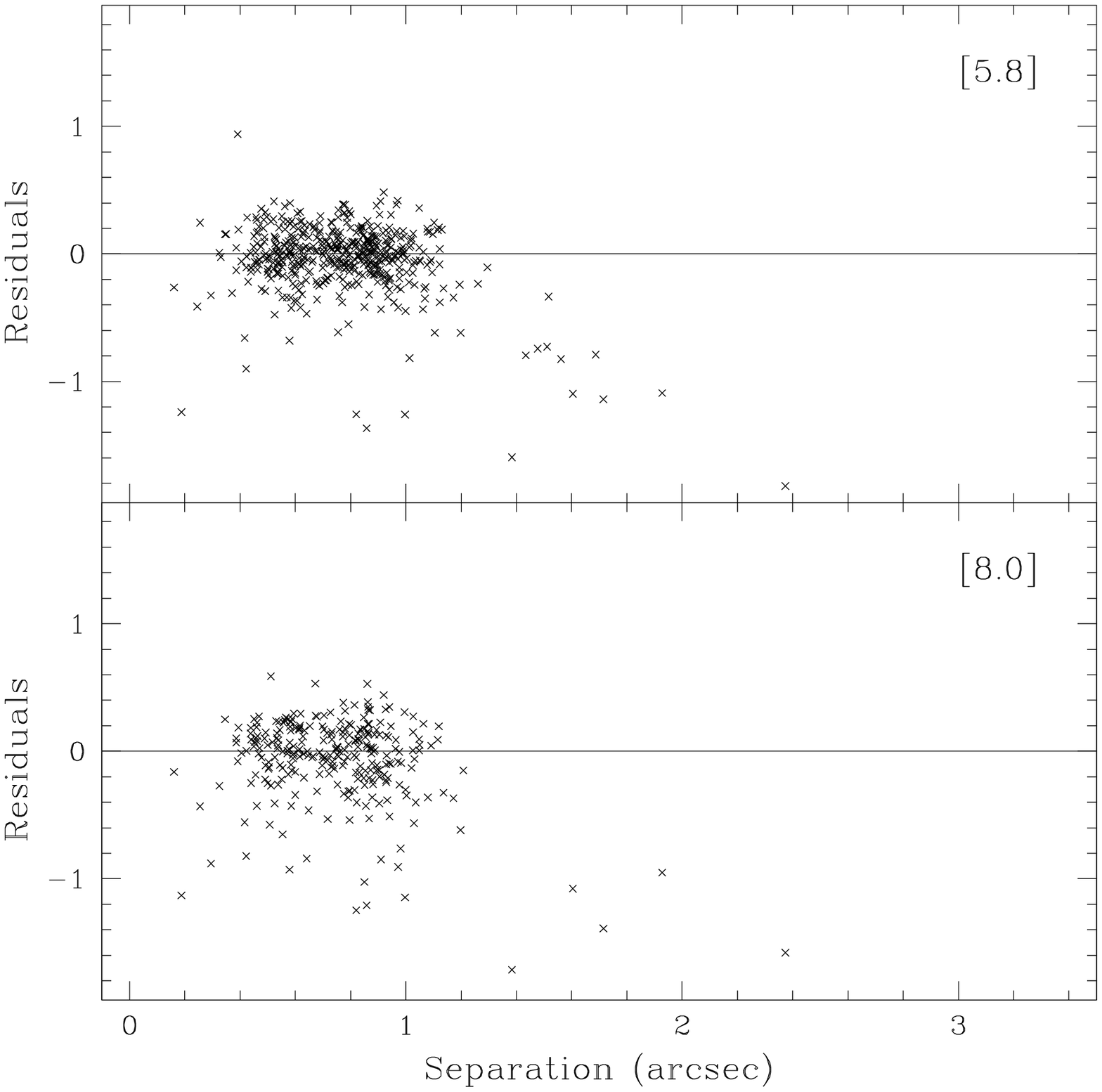}
\caption{The apparent magnitude residuals of the fits of the PL relations from \cite{Ngeow2008} as a function of the separation of the positions of OGLE II Cepheids and the infrared sources for (Left) $3.6$ and $4.5$ $\mu m$ and (Right) $5.8$ and $8.0$ $\mu m$.}
	\label{f1a}
	\end{center}
\end{figure*}

The are three possible causes of the infrared flux excess: (1) blending of stars in SAGE observations, (2) false matches of the infrared sources to the OGLE II Cepheids, (3) or circumstellar dust shells forming at a significant distance from the Cepheids in a stellar wind.  It is possible that false matches contaminate the sample, and we check this in Figure \ref{f1a} where the magnitude residuals of the IR Period--Luminosity relations determined by \cite{Ngeow2008} are shown as a function of the separation between the OGLE II and SAGE positions.  Although the search radius used in \cite{Ngeow2008} is rather large, most of the matched objects have a separation less than $0.77$ arcseconds [as demonstrated in the Figure 1 and Table 1  in \cite{Ngeow2008}].
The potential for false matches is only significant for a small fraction of the total sample of \cite{Ngeow2008}, and in the sample used here, only ten of the $488$ Cepheid matches have a separation greater than $1.3$ arcseconds, which is the separation where the matches almost all have large residuals to the fit of the PL relation.  There is another asymmetry apparent in Figure \ref{f1a} where the residuals have a separation less than $0.4$ arcseconds.  In this work, we keep the two samples, but identify them in the figures when important.

We can test if the infrared excess is due to mass loss by calculating the sum of the infrared luminosity of the Cepheid and the luminosity of the dust that is generated in the wind, given by
\begin{eqnarray}\label{eq1}
\nonumber&& L_{\nu, \rm{Shell}} = \frac{3}{4\pi}\frac{<a^2>}{<a^3>}\frac{Q_\nu^A }{\bar{\rho}}\frac{\dot{M}_d}{v_d}\\
&&\times \int_{R_*}^{\infty} B_\nu(T_d)[1 - W(r)]dr.
\end{eqnarray}
The dust shell luminosity is proportional to the ratio of the mean cross section, $<a^2>$, and volume, $<a^3>$, of the dust particles, and inversely proportional to the mass density of dust particles, $\bar{\rho}$.  The dust mass--loss rate is given by $\dot{M}_d$, the dust velocity is $v_d$, and $Q_\nu^A$ is the absorption efficiency.  The term in the integral is the product of the blackbody radiation of the dust with temperature $T_d$ and the geometric dilution factor $W(r) = [1 - \sqrt{1- (R_*/r)^2}]/2$ at a distance $r$ from the surface of the Cepheid. 

To predict the dust shell luminosity, we need to specify the properties of the dust.  The dust grain size is assumed to range from $0.005$ $\mu m$ to $0.25$ $\mu m$, which yields a value $<a^2>/<a^3> \approx 40$ $\mu m$ based on the method of \cite{Mathis1977}.  Assuming the dust is primarily graphite, the absorption will be concentrated at optical wavelengths, giving an absorption efficiency of $Q_\nu^A \approx 2$. This further implies that the mean density of the grains is $\bar{\rho} = 2.2$ $g/ cm^{3}$.  The dust velocity, equivalent to the terminal velocity of a wind, is about $100$ $km/s$, approximately equivalent to the escape velocity of a Cepheid.  The integral is computed from the surface of a Cepheid but dust does not form in the wind until the material is at a condensation distance $r_c =  (R_*/2)(T_*/1500K)^{5/2}$, where dust condenses at a temperature of $1500K$. The dust temperature at distance $r$ from the star, greater than the condensation distance, is $T_d(r) = T_*W(r)^{1/5}$. This leaves the dust mass--loss rate, stellar radius and effective temperature as unknowns in Equation \ref{eq1}.

The gas mass--loss rate is found by assuming a dust--to--gas ratio.  The typical ratio assumed for the Milky Way ISM is $1/100$, and this ratio was used in previous studies for mass--loss in Galactic Cepheids \citep{McAlary1986}.  The dust--to--gas ratio in the LMC must be significantly smaller than $1/100$ because the formation of dust depends on the metallicity of the gas.  For this work, a value of the LMC dust--to--gas ratio is assumed to be $1/250$ found by scaling the Milky Way dust--to--gas ratio by the ratio of the average LMC metallicity of $Z = 0.008$ to the standard solar metallicity $Z_\odot = 0.02$.  This choice of dust--to--gas ratio leads to a gas mass--loss rate that is a lower limit.  The dust--to--gas ratio in the LMC has been observed to be approximately one quarter the Galactic value  \citep{Clayton1985} to about one tenth the Galactic value \citep{Weingartner2001}.  Therefore a gas mass--loss rate may be smaller than would be predicted using other dust--to--gas ratios.  The dust velocity, which is also the terminal velocity of the gas wind, is chosen to be approximately the escape velocity, but the dust velocity may range from about $75$--$150$ $km/s$, leading to an uncertainty of about $50\%$. 

The mean luminosity of a Cepheid at frequency $\nu$ is given by
\begin{equation}\label{eq2}
L_{\nu,\rm{Star}} = 4\pi R^2_* \pi B_\nu(T_{\rm{eff}}),
\end{equation}
meaning the stellar luminosity is dependent on the radius and effective temperature.  We check if a blackbody is a reasonable approximation by comparing blackbody B, V, I brightnesses with the B,V, and I from a \textsc{Atlas} stellar atmosphere mode \cite{Kurucz1979} with an effective temperature of $6000K$ and $\log g = 1$.  The model atmosphere is consistent with Cepheid properties and the B,V, I agree with blackbody estimates to within a few tenths of a magnitude.  The infrared observations are well approximated by a blackbody brightness because these wavelengths are in the tail of the blackbody function at $6000$ $K$.   The different predicted brightnesses will affect the uncertainty of the mass--loss model but not greatly.  This leaves three unknown variables for fitting the observations: the dust mass--loss rate, the stellar radius and the effective temperature.  The effective temperature may be determined using the relation determined by \cite{Beaulieu2001}
\begin{equation}\label{eq3}
\log T_{\rm{eff}} = 3.930122 + 0.006776\log P - 0.2487(V-I)_0.
\end{equation}
The relation is dependent on the pulsation period and de--reddened color.

To compare the predicted total luminosity (stellar plus shell) of the LMC Cepheids with the observed fluxes, we need to adopt a distance modulus.  Our choice is $18.5 \pm 0.1$ for the LMC, which is approximately the mean distance modulus \cite[for example]{Catelan2008, Clement2008}, although the value ranges from $18.4$ to $18.7$. 

The errors for the optical and infrared observations are likely to be negligible compared to other uncertainties, so we assume that the error for each optical wavelength is $0.1$ magnitudes, due to the uncertainty of the distance to the Cepheids to determine the absolute magnitudes and the error for the infrared observations is $0.2$ magnitudes based on the distance uncertainty.  The main sources of error are the thickness of the LMC \citep{Lah2005}, error in the observations themselves and that the infrared observations may not be the mean brightness of  Cepheids.  In fact, a thickness of about $5\%$ of the distance to the LMC corresponds to an error of $0.1$ mag.

The BVI observations are also de--reddened while it is assumed that the extinction of the infrared light is negligible.  The color excess $E(B-V)$ is given in the OGLE II data for each Cepheid and is on average $\approx 0.15$.  The extinction for the BVI is calculated in the same way as in \cite{Udalski1999b}, while the extinction in the infrared is significantly less than $0.1$ magnitudes.

\section{Quality of Fit of Mass--Loss Rates}
\begin{figure*}[t]
	\begin{center}
	\epsscale{1.15}
		\plottwo{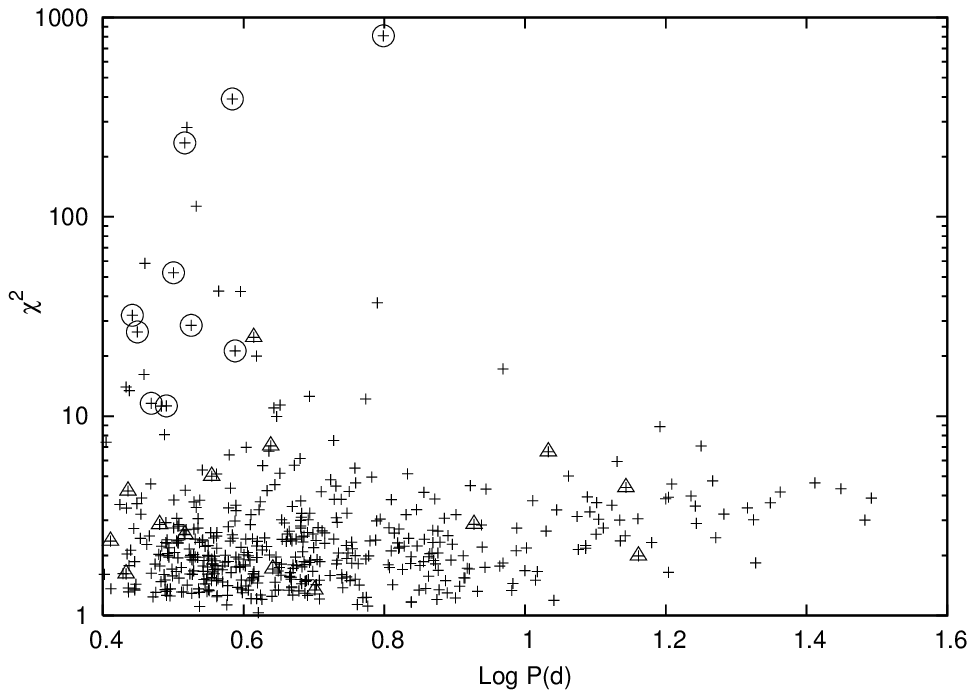}{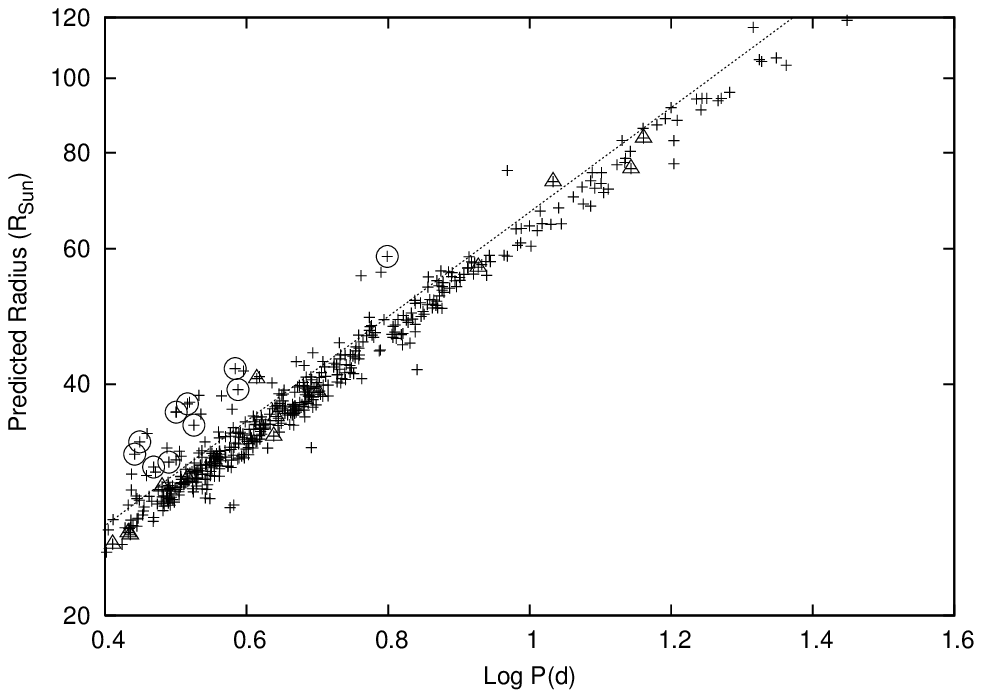}
\caption{(Left) The $\chi^2$ fit of the observations with best--fit mean radius only. (Right) The predicted radii of the sample of Cepheids with the Period--Radius relation \citep{Gieren1999} shown as a dotted line. The circled points are the Cepheids that are considered most likely to be false associations while the triangles represent the Cepheids with separation between OGLE II and SAGE coordinates of less than $0.4$ arcseconds.}
	\label{f1b}
	\end{center}
\end{figure*}

Having described the method for predicting stellar fluxes over the seven bands, we fit the mean radius and dust mass--loss rate, and hence the gas mass--loss rate, using $\chi^2$ fitting, for each Cepheid in the sample.  The value of $\chi^2$ is given as
\begin{equation}\label{eq4}
\chi^2 = \frac{1}{N-f}\sum_i^N \left(\frac{M_{\lambda_i}(\rm{Theory}) - M_{\lambda_i}(\rm{Obs})}{\sigma_i}\right)^2,
\end{equation}
where $N$ is the number of observations and $f$ is the number of unknowns.  The $\chi^2$ fits are calculated for a range of stellar radii based on the Period--Radius relation \citep{Gieren1999}
\begin{equation}\label{eq3a}
\log R/R_\odot = 0.68\log P(d) + 1.146.
\end{equation}
and varying that radius by $\pm 20\%$.   The dust mass--loss rate is fit for a specific mean radius by finding a minimum value of $\chi^2$ using a root--finding algorithm.  Before we attempt to fit the dust mass--loss rate, we first apply the $\chi^2$ fitting to the range of mean radii alone as a reference. The values of $\chi^2$ and predicted radii in this method are shown in Figure \ref{f1b}.

\begin{figure}[t]
	\begin{center}
	\epsscale{1.15}
		\plotone{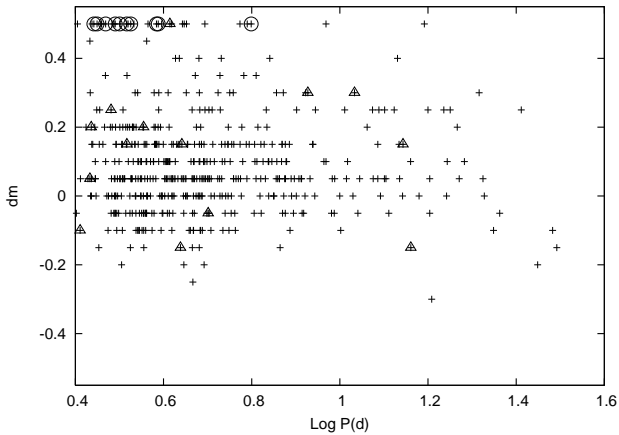}
\caption{The best--fit values of $dm$, the predicted differences between the mean brightness and observed brightness at infrared wavelengths.  Circled points are the likely false associations and triangles represent separations less $0.4$ arcseconds.}
	\label{f1c}
	\end{center}
\end{figure}

Fitting the radius alone to the observations seems to provide a reasonable fit, but one may argue that the uncertainty of the fit is related to the fact that the SAGE observations are single epoch and are not mean brightness at these wavelengths.  We test this by $\chi^2$ fitting the radius and a quantity $dm$.  This variable represents the difference between the observed infrared magnitude and the mean brightness of the Cepheids.  The values of $dm$ are assumed to vary from $-0.5$ to  $0.5$ magnitudes dimmer, where a negative value of $dm$ implies the mean brightness is brighter than what is observed.  The chosen range of values are exaggerated as the full brightness amplitude of a Cepheid at IRAC wavelengths is  $<0.4$ the amplitude in the visible.  The largest full amplitude in the visible is $1.2$ magnitudes, meaning the infrared variable $dm$ will be $ -0.25 < dm < 0.25$ in reality, and for the majority of Cepheids the range of $dm$ is much smaller.  We show in Figure \ref{f1c} the predicted values of $dm$ as a function of period.  If the hypothesized infrared excess were due primarily to the fact that the infrared observations are single epoch then about $50\%$ of the sample would predict $dm \le 0$ and $50\%$ $\ge 0$.  The results in Figure \ref{f1c} show a preference for $dm > 0$, with $68\%$ of the total sample having $dm > 0$ and $13\%$ preferring $dm = 0$.  Therefore, we take this as proof of an infrared excess in the SAGE sample.

\begin{figure*}[t]
	\begin{center}
	\epsscale{1.15}
		\plottwo{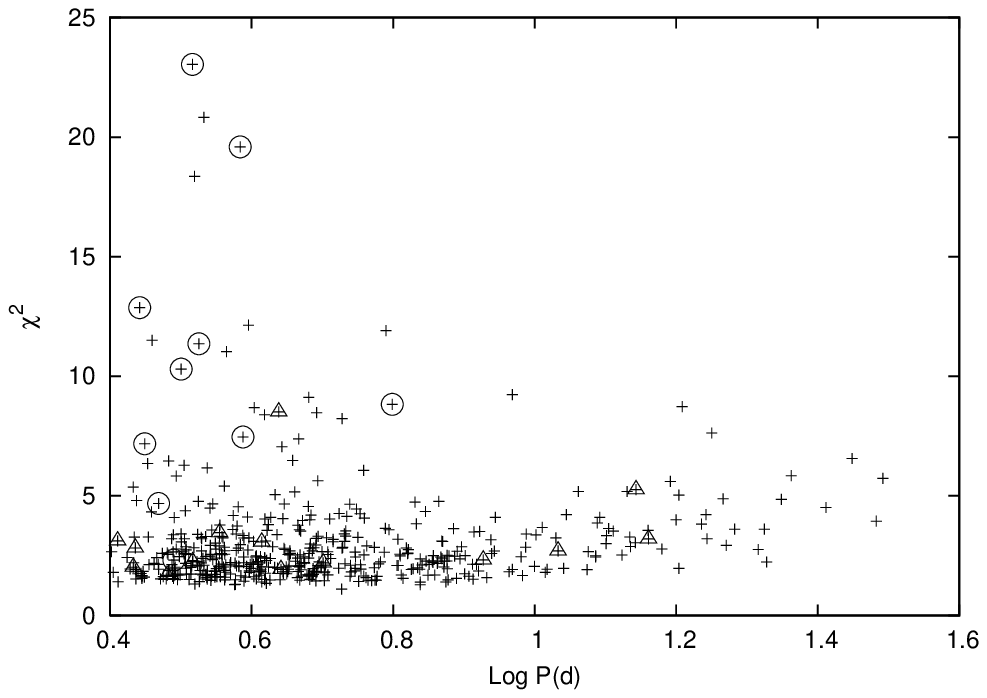}{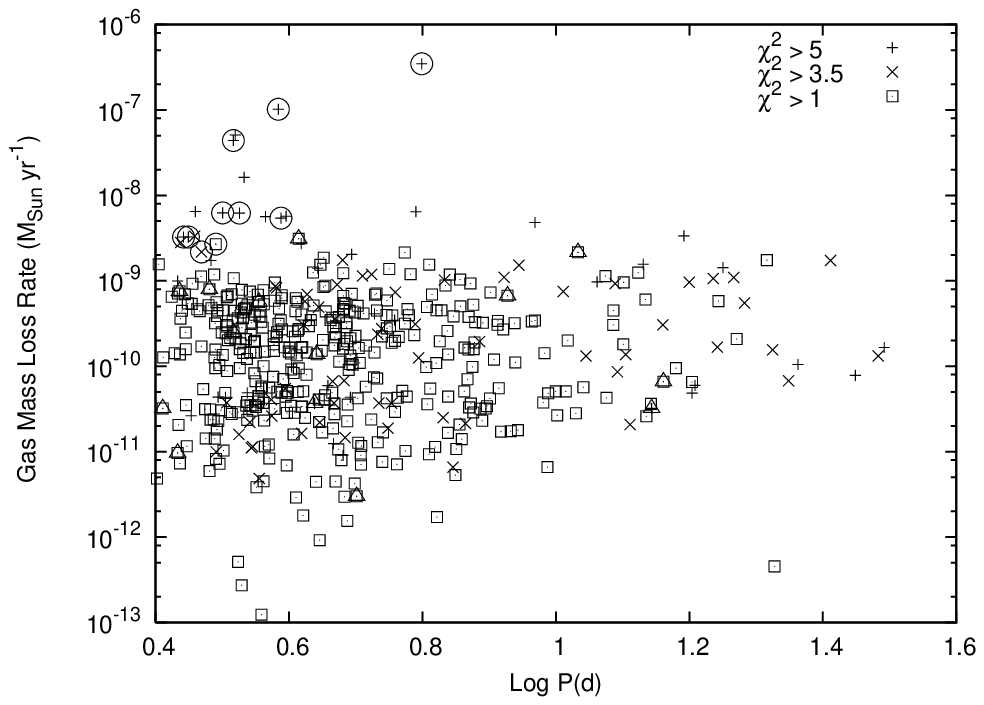}
\caption{(Left) The $\chi^2$ of the observation with best--fit mean stellar radius and mass--loss rate. (Right) The gas mass--loss rates of the sample of Cepheids, binned into different $\chi^2$ groups. Those points circled are the possible false associations.}
	\label{f2}
	\end{center}
\end{figure*}

The next step is to determine the best--fit radius and mass--loss rate for the sample of Cepheids. The $\chi^2$ fits are presented as a function of pulsation period in Figure \ref{f2} along with the gas mass--loss rate.
The mass--loss rates range from $10^{-12}$ to $10^{-8}M_\odot/yr$, with the values of $\chi^2$ ranging from about $1.1$ to about $23$.  The large majority of Cepheids appear well fit by a mass--loss model that forms dust at some distance from the surface of the Cepheid.     These predicted mass--loss rates are significant, $~10^{-9}M_\odot/yr$, and, depending on the dust--to--gas ratio and dust grain properties, the gas mass--loss rates may be an order of magnitude larger or even more.  These gas mass--loss rates are the minimum mass--loss rates for the LMC Cepheids.  

\begin{figure}[t]
	\begin{center}
	\epsscale{1.15}
		\plotone{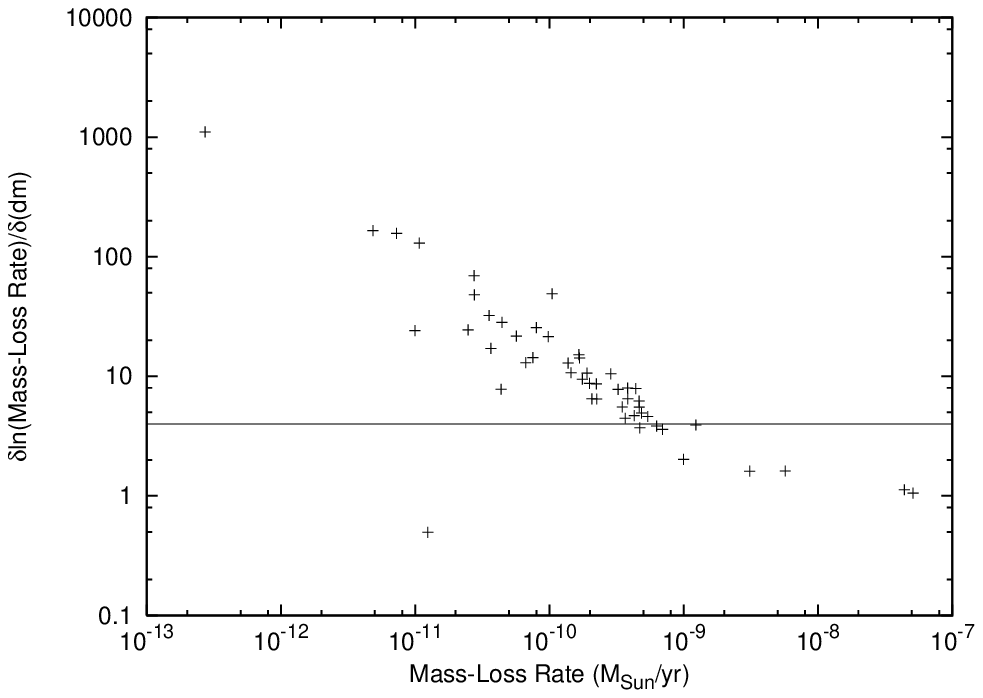}
\caption{The uncertainty of the mass--loss rates due to the difference between the observed infrared brightness and the unknown mean brightness of the Cepheids as a function of mass--loss rate. The horizontal line refers to an uncertainty of $100\%$ for a full pulsation amplitude of $0.5$ mag.}
	\label{f2b}
	\end{center}
\end{figure}

We also quantify the uncertainty of the mass--loss rates caused by the unknown phase of the infrared observations by computing the best $\chi^2$ fits for the mass--loss rate, radius and the quantity $dm$ that was defined earlier, and we compute the values of $\delta \ln(\dot{M})/\delta (dm)$ for a random, with respect to period, subsample of 100 of the Cepheids.  This error is a function of both the pulsation amplitude, and the mass--loss rate.  We show the values of $\delta \ln(\dot{M})/\delta (dm)$ as a function of $\dot{M}$ in Figure \ref{f2b}.  The uncertainty is related to the mass--loss rate, and, as one would expect, the uncertainty of the pulsation amplitude is  less important for larger predicted mass--loss rates.  We highlight the boundary where the uncertainty of the mass--loss rate is $100\%$ for a pulsation amplitude of $0.5$ magnitudes which represents the maximum infrared pulsation amplitude. This implies that Cepheids with predicted mass--loss rates $> 10^{-9}M_\odot/yr$ have infrared excesses that cannot be explained solely by pulsation phase, and a number of Cepheids with mass--loss rates $<10^{-9}M_\odot/yr$ have uncertainties $\delta \ln(\dot{M})/\delta (dm) < 4$.  As the IR pulsation amplitudes become known we will be able to probe smaller mass--loss rates with more certainty.

\begin{figure*}[t]
	\begin{center}
	\epsscale{1.15}
		\plottwo{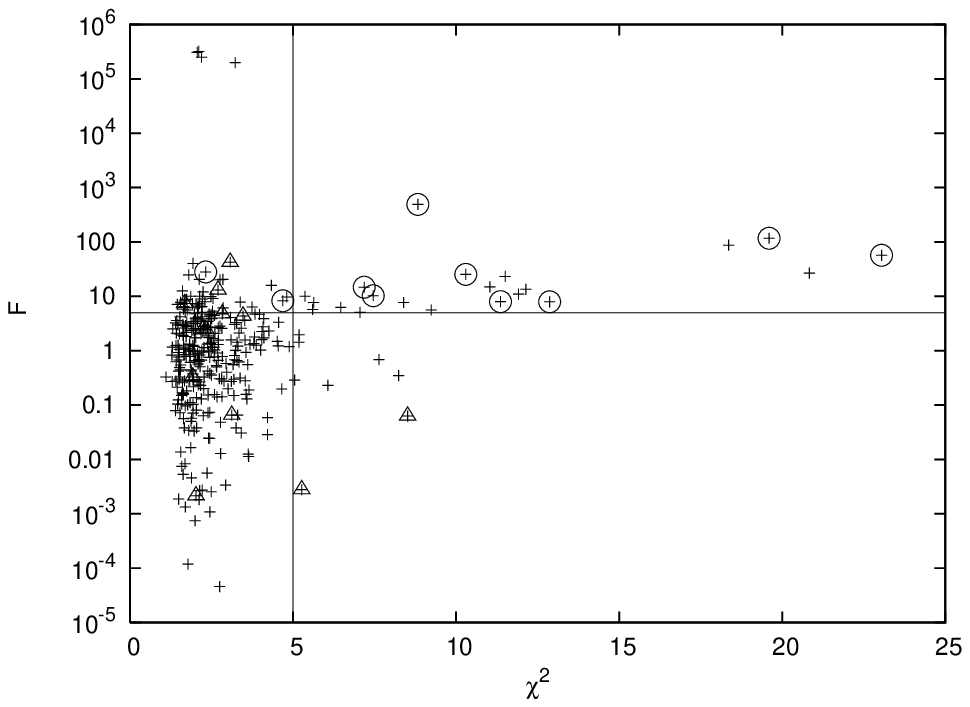}{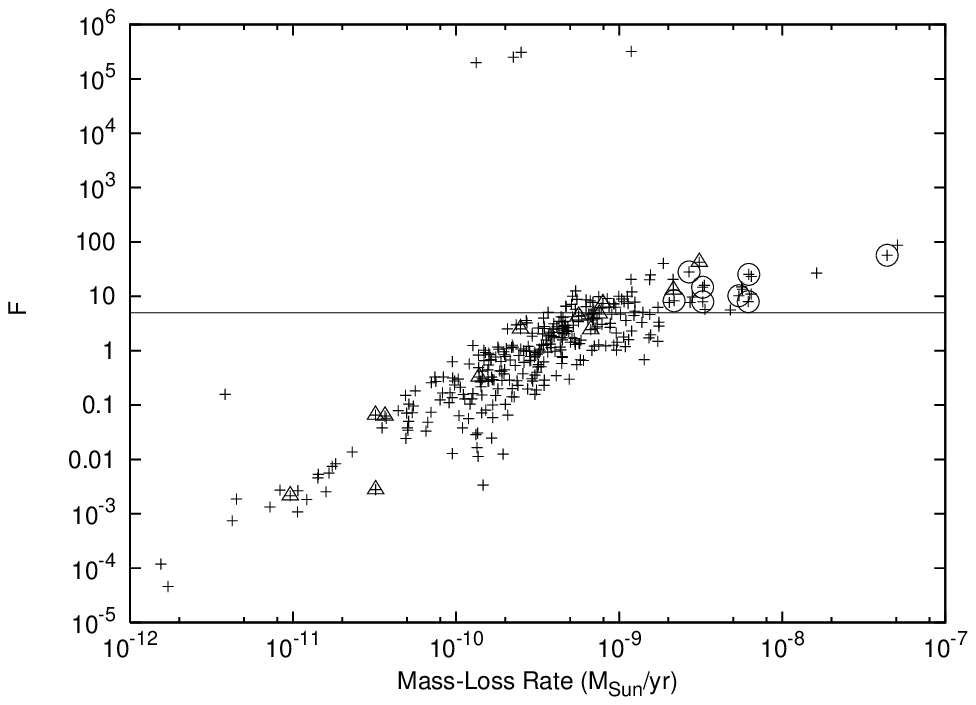}
\caption{(Left) Calculation of the F--test for each Cepheid in the sample plotted in terms of the value of $\chi^2$ from the mass--loss model. The horizontal line represents the $95\%$ confidence level that the mass--loss model is distinguished from fitting the mean radius only.  The vertical line is a represents a cut--off of $\chi^2 = 5$ where we interpret $\chi^2 < 5$ as a reasonable fit. (Right) The values of F as a function of the predicted mass--loss rates.  The circled points represent those Cepheids that are likely false associations of the infrared observations and the triangles represent those with separations less than $0.4$ arcseconds.}
	\label{f3}
	\end{center}
\end{figure*}

The model is also tested by comparing the two parameter fits, with the mass--loss rate and radius as the two degrees of freedom, to fits using just the radius as the only degree of freedom.  We use the F--test to quantify the significance of the mass--loss model. The F--test is described by \citet[and references therein]{Kanbur2004,Ngeow2008} and for each Cepheid we calculate the value of F.  For the majority of Cepheids, the fit of the radius has only six degrees of freedom while the mass--loss model has five.  Some of the Cepheids have one less degree of freedom for each model respectively.  The values of F are shown in Figure \ref{f3} against the values of $\chi^2$ for the mass--loss model.  
The values of $F > 5$ mean that we can state with  $95\%$ confidence that the mass--loss model is significant relative to fitting the radius alone. This provides no information into the possibility of blending, for instance, so we also take a cut of models with a value of $\chi^2 < 5$.  This corresponds to the upper left quadrant of Figure \ref{f3} which contains 44 Cepheids.  Therefore we state with $95\%$ confidence that  approximately $9\%$ of the sample of LMC Cepheids have circumstellar dust shells caused by stellar winds.

It is shown that 44 of the Cepheids are consistent with the mass--loss model, implying that the remainder of the sample is consistent with no mass loss.  However the value of $F$ is a function of the mass--loss rate, with $F$ increasing with $\dot{M}$, as shown in Figure \ref{f3}.  Cepheids with values of $F<9$ and $\chi^2 < 5$ in the lower left quadrant of Figure \ref{f3} with $\chi^2 < 5$ are consistent with mass loss but the observational errors are too large to state with certainty that all LMC Cepheids are undergoing mass loss.  This suggests that the model needs to be tested with time series infrared observations to constrain the pulsation amplitude and reduce the uncertainty of the infrared observations.  In the next section, we test what effect mass loss might have on the infrared PL relation if the hypothesis is correct.

\section{The Effect of Mass Loss on Infrared Period Luminosity Relations}
It has been postulated that mass loss generates dust in a circumstellar shell surrounding a Cepheid and this, in turn, affects the infrared PL relation.  By using the predicted stellar luminosities of the sample of Cepheids, we compute the stellar PL relations, $m_\lambda = a\log P + b$, and compare them with results from \cite{Ngeow2008} and \cite{Freedman2008}.  We also test the data for non--linearity in the infrared PL relations.  In the fit, we do not include the Cepheids with large separations that were noted in the previous section, however, we do include those with separation less than $0.4$ arcseconds because they have randomly distributed mass--loss rates, in Figure \ref{f2},  and hence have randomly distributed infrared excesses.

\begin{figure*}[t]
	\begin{center}
	\epsscale{1.15}
		\plottwo{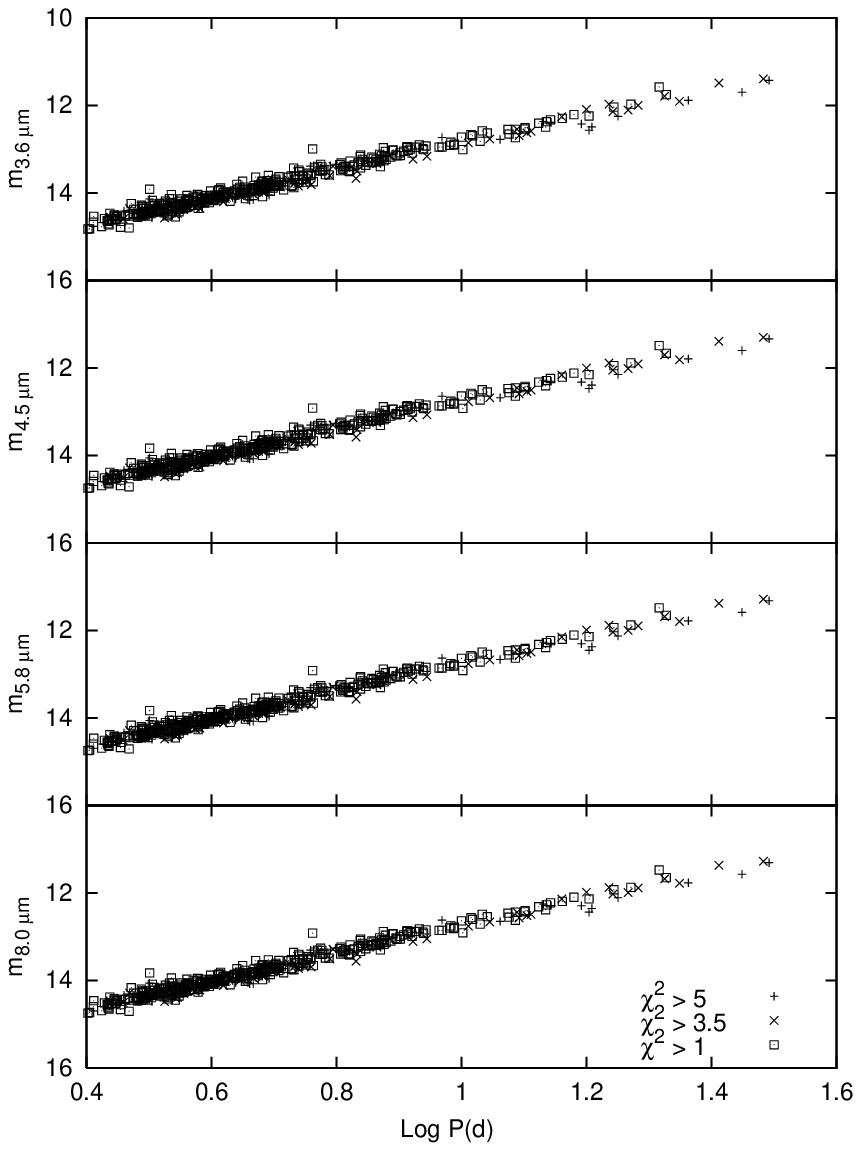}{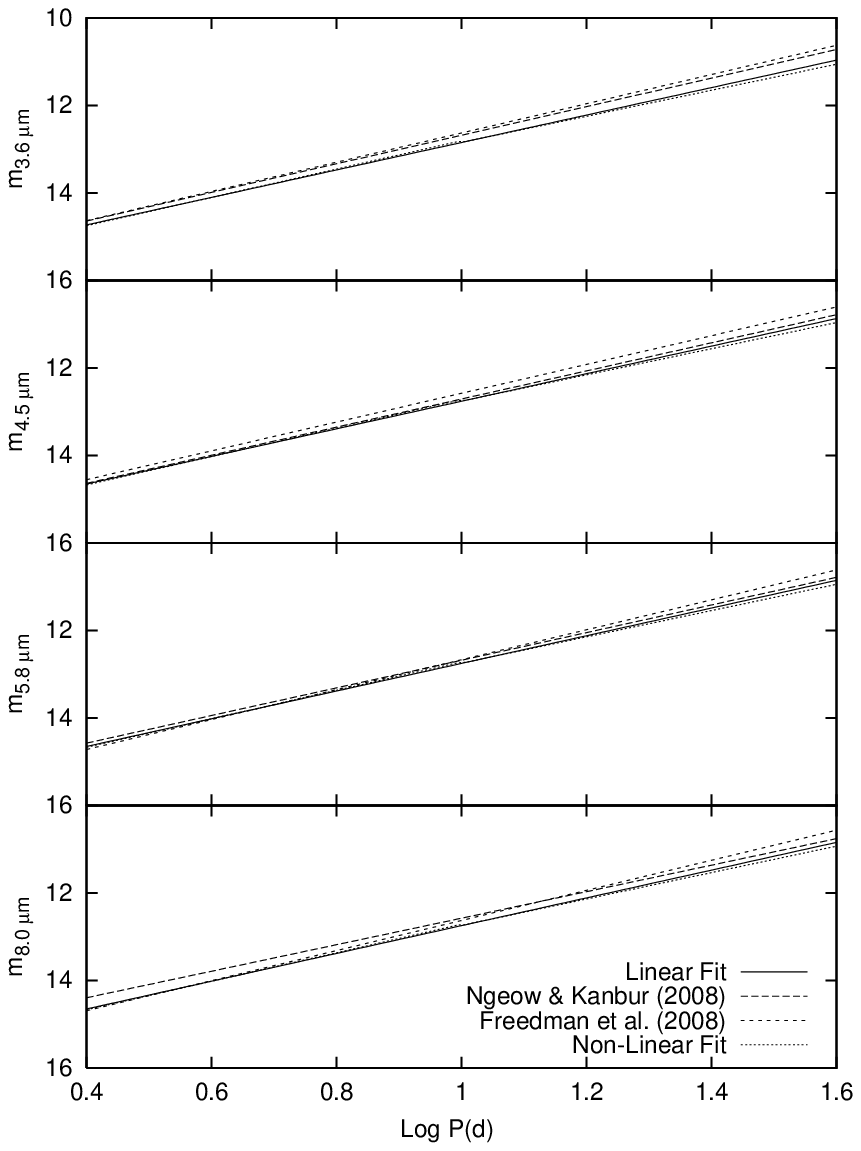}
\caption{(Left) The predicted stellar brightnesses of the LMC Cepheids as a function of period.  (Right) The Comparison of the best--fit linear and non--linear PL relations to the relations determined by \cite{Ngeow2008} and \cite{Freedman2008} where we removed Cepheids with separation greater than $1.3$ arcseconds.}
	\label{f4}
	\end{center}
\end{figure*}

The predicted stellar luminosities of the LMC Cepheids in the mass-loss model are shown in Figure \ref{f4}, together with a comparison of the linear and non--linear PL relations with the relations from \cite{Ngeow2008} and \cite{Freedman2008}. The non--linear PL relation is defined as two linear relations, the first for the period range of $1$ to $10$ days while the second is for the longer period range \citep{Ngeow2005}. The slopes, zero--points and dispersions of the fits for the linear and non--linear fits are given in Table \ref{t1}.   

\begin{table}[t]
\caption{Best Fit Parameters for Predicted PL Relations}
\begin{center}
\begin{tabular}{lcccc}
\hline
Type & $\lambda$ $(\mu m)$ & Slope & Zero Point & Dispersion \\
\hline
Linear & $3.6$ & $-3.145 \pm 0.024$ & $15.993 \pm 0.017$ & $0.110$ \\
 &$4.5$ & $-3.159\pm 0.023$ & $15.921 \pm 0.017$ & $0.108$\\
&$5.8$ & $-3.170\pm 0.023$ & $15.924\pm 0.017$ & $0.107$ \\
& $8.0$ & $-3.181 \pm 0.023$ & $15.929 \pm 0.017$ & $0.105$ \\
\hline
Non--  & $3.6$ &$-3.248 \pm 0.038$ & $16.057 \pm 0.025$ & $0.107$ \\
 Linear & $4.5$ &  $-3.259 \pm 0.038$ & $15.983\pm 0.025$ & $0.105$ \\
$P <10d$ & $5.8$ & $-3.268 \pm 0.037$ & $15.984 \pm 0.025$ & $0.104$ \\
 & $8.0$ & $-3.276 \pm 0.037$ & $15.988 \pm 0.024$ & $0.103$ \\
 \hline
 Non-- & $3.6$ & $-2.971 \pm 0.123$ & $15.815 \pm 0.146$ & $0.125$ \\
Linear  & $4.5$ & $-2.989 \pm 0.121$ & $15.747 \pm 0.143$ & $0.122 $\\
$P>10d$& $5.8$& $-3.005 \pm 0.119$ & $15.754 \pm 0.141$ & $0.120 $\\
 & $8.0$ & $-3.019 \pm 0.118$ & $15.763 \pm 0.140$ & $0.118$ \\
\hline
\end{tabular}
\end{center}
\label{t1}
\end{table}

The linear PL relations have smaller dispersion than the PL relations from \cite{Ngeow2008} for the two longer wavelengths and slightly larger dispersion for the two shorter infrared wavelengths.  There are also notable differences in the slopes and the zero points.  The zero points of the predicted $3.6$ and $4.5$ $\mu m$ PL relations are approximately the same as those determined by \cite{Ngeow2008}, while at longer wavelengths the differences are significant and also the predicted zero points tend to be a little brighter than the zero--points found by \cite{Freedman2008}. This is consistent with mass loss causing larger infrared excesses at longer wavelengths.  The predicted slopes range from $-3.14$ at shorter wavelengths to $-3.18$ at longer wavelengths.  This is a small change of slope as a function of wavelength, and it is roughly consistent with a constant slope within the errors given n Table \ref{t1}.  The slopes from \cite{Ngeow2008} show the opposite behavior  with the slopes becoming less steep with longer wavelength, contrary to the arguments in \cite{Freedman2008} who maintain the slope of the PL relation should be steeper as a function of wavelength and approaches an asymptotic limit.  In both cases the uncertainty of the slope is similar, ranging from $0.017$ to $0.048$ with increasing wavelength for \cite{Ngeow2008}, and about an average of $0.03$ for \cite{Freedman2008}.  Therefore the slopes from these two works and those predicted here do not agree within the uncertainty.

This shift in the behavior of the slopes of the infrared PL relations is due the removal of infrared excess caused by mass loss.  This is best seen at $8.0$ $\mu m$ where the slope changes from about $-3$ \citep{Ngeow2008} to $-3.18$ when the contribution of brightness due to mass loss is removed, although the observed slope of $-3$ may also be due to incompleteness of the data at the faint end of the PL relation.  The  implication is that mass loss in short period Cepheids contributes significant luminosity, increasing the zero point and causing a shallower slope because the majority of the Cepheids in the sample have periods less than ten days.  The shallower slope implies that mass loss  contributes fractionally less to the total infrared luminosity at longer periods because the stellar luminosity is already so large.

\begin{figure}[t]
	\begin{center}
	\epsscale{1.15}
		\plotone{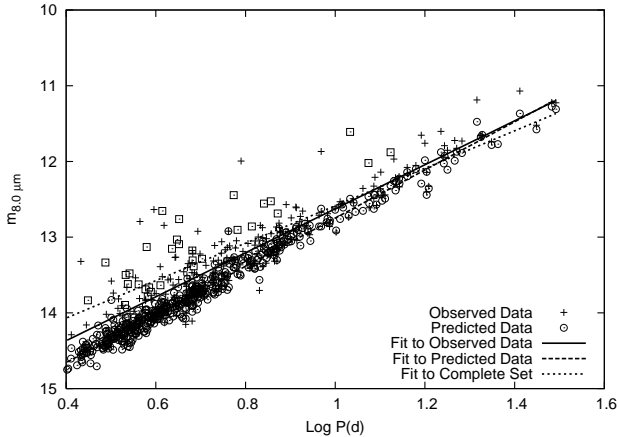}
\caption{Comparison of the predicted stellar and the observed fluxes of the sample of Cepheids.  The 44 Cepheids where the mass--loss model is statistically unique are shown as squares.  The lines represent the predicted stellar flux, the observed stellar flux of the sample and that of the complete set from \cite{Ngeow2008}.}
	\label{f4a}
	\end{center}
\end{figure}

The analysis in the previous section showed that the mass loss hypothesis is statistically unique from fitting only the radius of the Cepheid for about $44$ Cepheids in the sample, or conversely that the majority of Cepheids in the sample are statistically consistent with zero mass loss.  From this realization, we wish to compare the observed infrared brightness that we fit our model with and the predicted infrared brightness of the Cepheids.  The comparison is shown in Figure \ref{f4a} for the $8.0 $ $\mu m$ data.  At this wavelength the differences is most apparent because a dust shell contributes a larger fraction of the total flux at longer wavelengths. We also compute best--fit linear relations for the observed data where the Cepheids with large separation and the $44$ Cepheids where mass loss is shown to be likely are not used in the fitting. This relation is  $m_{8.0\mu m}(\rm{Observed}) = -2.905\log P + 15.530$ with a standard deviation of $0.219$.  For comparison, we derive the best--fit data using all of the $8.0$ $\mu m$ data from \cite{Ngeow2008} and find $m_{8.0\mu m}(\mbox{Complete}) = -2.473\log P + 15.058$ with a standard deviation of $0.602$.  The relations given in \cite{Ngeow2008} are determined using an iterative fitting method where a best--fit relation is determined and then any Cepheids with a brightness that is more than $3\sigma$ different are removed and a new PL relation is computed and the process repeats until the PL relation converges. Here, we compute the PL relation using all of the data without the iterative approach,  The linear relations are shown in Figure \ref{f4a}.  Although we are only able to confidently state that 44 of the Cepheids are consistent with the mass--loss model, we note that there are significant differences between the predicted stellar brightnesses and observed data, and that these differences are reflected in the infrared PL relations.  It is also interesting that these 44 Cepheids are observed to be brighter than the majority of the sample but there are a number of Cepheids with similar brightness that are statistically consistent with zero mass loss.

The data are tested for non--linearity in the infrared PL relations.    The hypothesis that the infrared PL relations are non--linear is tested with the F--test, as described in \citet[and references therein]{Kanbur2004,Ngeow2008} by comparing a PL relation of the form
\begin{equation}
m_\lambda = \left\{\begin{array}{c} a\log P + b \mbox{\hspace{1cm}} \log P < 1 \\
 c \log P + d \mbox{\hspace{1cm}} \log P > 1\end{array}\right.
\end{equation}
with the standard linear PL relation with two degrees of freedom. If the value of $F>3$, the PL relations are non--linear with $95\%$ confidence.  Our values of $F$, with increasing wavelength, are $7.89, 5.93, 5.78$, and $5.50$.  The predicted stellar PL relations are thus consistent with being non--linear with a period break at $10$ days. The predicted slopes and zero--points are given in Table \ref{t1}.   However there are two possible sources of error.   The first is that we are assuming blackbody radiation that ignores any infrared absorption lines that may affect the structure of the PL relations.   The second is that there are significantly less data for periods greater than $10$ days (approximately $50$ data points). 

The non--linearity is related to the fact that the luminosities are given by the effective temperature;  using the OGLE II data to derive effective temperatures may cause a non--linear Period--Temperature relation because of non--linearity in the OGLE II $(V-I)$ Period--Color relation \citep{Kanbur2004a}.  This non--linear Period--Temperature relation causes non--linearity in the infrared predictions.  This implies that the PL relations given by only the stellar component is non--linear in the wavelength range of $3.6$ to $8.0$ $\mu m$ based on blackbody arguments, contradicting the results of \cite{Ngeow2008} and \cite{Ngeow2006} for the K--band PL relation. There are two plausible reasons why this contradiction is found. \cite{Kanbur2004b} argue the non--linearity is due to the hydrogen ionization front (HIF) interacting with the photosphere, causing significant temperature variations in the layers of the Cepheids that emit mostly in the optical; at longer wavelengths this interaction becomes less significant.  This implies that the mean effective temperature at shorter periods is affected by the HIF while at longer periods the effective temperature is just what would be expected for a non--pulsating star.  At infrared wavelengths, most of the radiation is emitted higher in the stellar atmosphere farther from the effects of the HIF, which leads to a more linear PL relation and is hence more dependent on the Period--Radius relation, which is linear.  This would explain why the values of $F$ for the non-linear relations decrease with longer wavelength, the IR PL relations are becoming more consistent with a surface brightness related to the linear Period--Radius relation.  The non--linearity of the predicted data may just be reflecting the non--linearity in the optical wavelengths because the variations of the effective temperature over the pulsation period is ignored. 

The second possibility is that mass loss causes larger infrared excess for shorter period Cepheids than for longer period Cepheids even though the mass--loss rates are similar for short ($P<10d$) and long ($P > 10d$) period Cepheids.  The short period Cepheids have smaller radii leading to smaller, more dense circumstellar shells.  The more dense shells cause greater infrared excess.  This greater infrared excess in shorter period Cepheids makes them appear brighter on average, which increases the zero point of the infrared PL relation; because the relative infrared excess decreases with longer period, the slope of the PL relation will appear shallower, with the effect being more prominent at longer wavelength.  This idea may explain the marginal linearity found in the K--band PL relation.  Infrared excess in Galactic Cepheids has been observed using K--band interferometry \citep{Kervella2006,Merand2006, Merand2007} so it is likely infrared excess plays a role in the LMC Cepheids at this wavelength.  This would imply that the K--band PL relation is actually non--linear and this non--linearity is being masked by the infrared excess.  This argument also explains the results of the tests of non--linearity in the IRAC PL relations in \cite{Ngeow2008}, in particular the non--linear PL relation at $8.0$ $ \mu m$.  The authors found that the slope of the non--linear PL relation for $P<10d$ is shallower than the slope of the linear relation at $8.0$ $\mu m$ with a more luminous zero point.  The non--linear PL relations for $P < 10d$ in the optical and near--IR all display the opposite behavior with respect to the linear PL relations.  This non--linear relation at $8.0$ $\mu m$ is due to the same process that causes the other IRAC PL relations to appear linear except the process is more significant at longer wavelengths.

It has been shown that mass loss provides a significant contribution to the infrared brightness of LMC Cepheids and affects the structure of infrared PL relations.   Without the contribution of mass loss, the slopes of the linear PL relations are steeper with increasing wavelength albeit at a small rate differing from the slopes becoming more shallow as found by \cite{Ngeow2008}.  Applying the F--test to the predicted data implies that the infrared PL relations are non--linear, though this result requires further testing.  However most of the Cepheids have predicted mass--loss rates that are statistically consistent with zero implying this result is preliminary and needs to be tested further with more data with smaller uncertainties.

\section{What is the Driving Mechanism?}
Up to this point, we have investigated the ability of mass loss to match the OGLE II and SAGE observations of LMC Cepheids, and how the resulting estimates of infrared excess affects the structure of the PL relations.  This has been done \emph{without} assuming a driving mechanism of the Cepheid wind. There are a number of possible methods for stars to drive mass loss, but only two are likely for Cepheids: radiative driving and pulsation driving. The arguments for these two possibilities are given in \cite{Neilson2008}, who also derive a model for pulsation--driving in Cepheids.  It is not feasible to apply the pulsation--driving model to this set of data as we do not have knowledge of the pulsation amplitudes or masses to which the model is sensitive.  However we can test whether a \emph{radiative}--driven stellar wind can match the predicted mass--loss rates using the method of \cite*{Castor1975}.  

\begin{figure*}[t]
	\begin{center}
	\epsscale{1.15}
		\plottwo{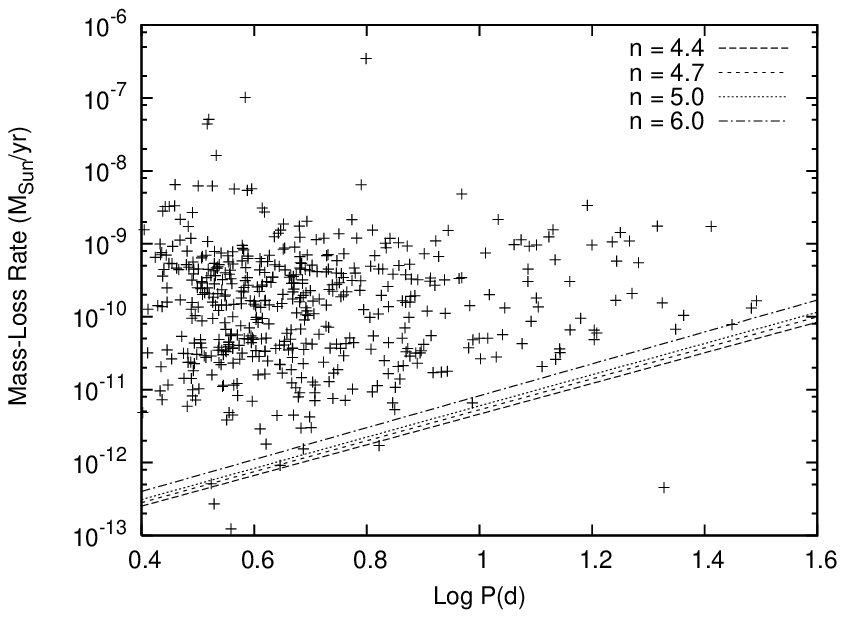}{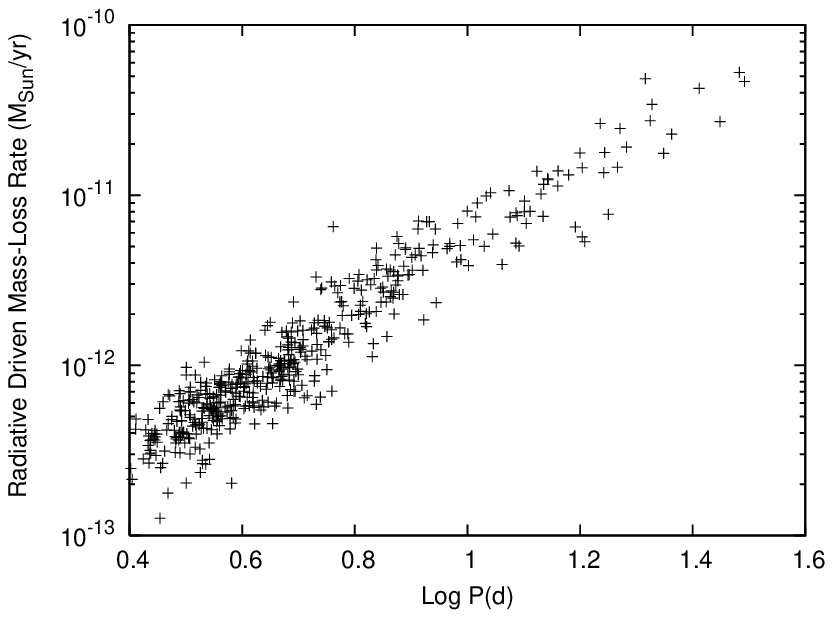}
\caption{(Left) The comparison of the mass--loss rates found from the infrared observations with best--fit linear relations representing mass--loss rates found from radiative--driving calculations, and  (Right) the radiative--driven mass--loss rates for the mass--luminosity relation where $n = 4.7$.}
	\label{f5}
	\end{center}
\end{figure*}

The calculation of the mass--loss rate for a radiative--driven stellar wind is reviewed in \cite{Lamers1999}, and \cite{Neilson2008} and will not be repeated here.  To conduct the calculation the mass, luminosity, radius and effective temperature are needed; the radii are determined by the $\chi^2$ fitting, the effective temperatures are given by the relation from \cite{Beaulieu2001} and the luminosity is found from the radius and effective temperature.  The mass is unknown so the radiative--driven mass--loss rates are found using a number of masses via mass--luminosity relations from \cite{Bono2000}, where $L = M^n$ in solar units.   The mass--loss rates found from the observations are shown in the left panel of Figure \ref{f5}, plotted with the best--fit relations for radiative--driven mass--loss rates found with the following mass--luminosity relations, where $n = 4.4,4.7, 5.0$, and $6.0$. The value of $n = 4.4$ represents the mass--luminosity relation from stellar evolution calculations, while $n = 4.7$ and $5.0$ represent the mass--luminosity relations relating to mass found using pulsation calculations, and $n = 6.0$ is used as an extreme case.   An example of the values of the radiative--driven mass--loss rates is shown in the right panel of Figure \ref{f5} for the case of $n = 4.7$.  Radiative--driven mass--loss rates for other values of $n >4.7 $ will increase the rates and for $n<4.7$ will decrease the rates by a roughly constant amount for each Cepheid.

The radiative--driven mass--loss rates are significantly smaller than the mass--loss rates determined from the observations.  At short periods of approximately $5$ days, the radiative mass--loss rates are about $10^3$ to $10^5$ times lower.  However, at periods greater than $30$ days the radiative--driven mass--loss rates are of similar order as the calculations.  This implies that the mass--loss cannot be driven by radiative lines alone at short period; there must be another driving mechanism.    This differs at longer period, but it should be noted that the mass--loss rates found from the observations are the minimum value based on the dust--to--gas ratio.  This means that the predicted gas mass--loss rates from infrared observations may be larger than the \emph{radiative}--driven mass--loss rates. 

As a further test of whether the mass loss is consistent with radiative driving we compute the circumstellar flux from dust created in a radiative--driven wind and added that to the predicted blackbody fluxes to compute infrared PL relations.  These relations are predictions of what would be observed if the mass loss is consistent with radiative driving.  The fitted relations at the four wavelengths $3.6,$ $4.5$, $5.8$, and $8.0$ $\mu m$ have slopes and y--intercepts that are equivalent to the IR PL relations derived from the predicted stellar fluxes alone  within the error of the fits.  For instance the slope and y--intercept of the $8.0$ $\mu m$ relation is $-3.139 \pm 0.024$ and $15.988 \pm 0.017$, differing by only a few thousandths from the $8.0$ $\mu m$ PL relation determined from the predicted stellar fluxes alone.  Radiative driving does not explain the significant infrared excess of 44 Cepheids that are explained by mass loss.

The amount of mass loss from the LMC Cepheids do not agree with radiative--driving calculations, and the mass loss is more consistent with the pulsation--driven model of \cite{Neilson2008, Neilson2008b} if one considers the magnitude of the mass--loss rates and the amount of scatter, especially for the 44 Cepheids where the mass--loss model is unique.  The remainder of the sample cannot be distinguished statistically from pulsation driving, radiative driving or no mass loss.  However, if one invokes no mass loss or radiative driving then it is more difficult to explain the behavior of the 44 Cepheids.  Pulsation--driven mass loss predicts that mass loss is driven by shocks generated in a pulsating atmosphere and the shocks tend to be more efficient at hotter effective temperatures.  This would explain the large range of mass--loss rates at similar pulsation periods.  The pulsation--driven mass--loss rates tend to be orders of magnitude larger than radiative--driven mass--loss rates as shown for observations of Galactic Cepheids and theoretical models of Galactic, LMC and SMC Cepheids at shorter periods.  We conclude that the mass--loss rates found in this work provide evidence for the model of pulsation--driven mass loss.

\section{Discussion and Conclusions}
In this article, we hypothesized that LMC Cepheids have significant infrared excess based on infrared observations of the LMC from the SAGE survey, and that the infrared excess is caused by dust forming in a Cepheid wind at a large distance from the surface of the star.  The idea was tested using OGLE II BVI observations along with IRAC observations of the Cepheids to best--fit radii and dust mass--loss rates of the LMC Cepheids.  The effective temperatures are determined using a temperature--color--period relation from \cite{Beaulieu2001}.  

The predicted gas mass--loss rates are significant with an average about $10^{-10}$ to $10^{-9}M_\odot/yr$ and may possibly be as high as $10^{-7}M_\odot/yr$, depending on the value of the dust--to--gas ratio. These mass--loss rates are not consistent with have radiative line--driving as the primary driving mechanism for the LMC Cepheids.  The rates, instead, provide evidence for shocks and pulsation driving the mass loss when compared to the analytic model from \cite{Neilson2008,Neilson2008b}.

The mass--loss model is compared to the fit of the observations with radii alone, and it is shown that a model fitting only the radius may be rejected with $95\%$ confidence relative to the mass--loss model for 44 Cepheids but this means that the remainder of the sample is consistent with no mass loss.  Therefore the results and arguments in the work based on the mass--loss model should be regarded with caution.  For almost every case the predictions are limited by the uncertainties of the distance modulus and the IR pulsation amplitude.  This comparison of the two models using the F--test is dependent on the mass--loss rate, and this shows we detect reliable infrared excess if the mass--loss rate is $> 10^{-10} M_\odot/yr$. Because of the dependence of the F--test on the mass--loss rate we explored what effect mass loss would have on the infrared PL relation.  The mass--loss model would benefit from infrared observations over the period of pulsation to determine the  mean brightness, which would decrease the uncertainty of the infrared brightnesses. 

The large mass--loss rates of LMC Cepheids may explain the Cepheid mass discrepancy in the LMC.  The pulsation masses tend to be about $20\%$ smaller than evolution masses in the LMC \citep{Keller2008}, which translates to a difference of about $1M_\odot$ for lowest mass Cepheids up to a few solar masses for the most massive Cepheids.  The mass--loss rates found in this work agree with the discrepancy for low mass Cepheids to an order of magnitude because the evolutionary timescale for a Cepheid on its second crossing is of order ten million years.  However the mass--loss rates are too small to be consistent with a $20\%$ mass discrepancy for the more massive Cepheids.  It should be noted that the mass discrepancy in LMC Cepheids is measured from Cepheids with periods less than $20d$, which have evolutionary masses from about $4$ to $7M_\odot$ \citep{Bono2000}.  The mass discrepancy has not been measured for more massive LMC Cepheids.

It has also been found that mass loss affects the infrared PL relations.  Using the predicted stellar luminosities we constructed new infrared PL relations that do not have infrared excess. These relations differ from those determined by \cite{Ngeow2008} with differences in the zero point and the slope.  The IR PL relations from \cite{Ngeow2008} have slopes that become smaller at longer wavelength inconsistent with the argument that the slope of the PL relation should approach a constant maximum value at longer wavelength base on the Period--Radius relation \citep{Freedman2008}.  The slopes in this work are all about $-3.15$ with a small amount of steepening at longer wavelength.  This would imply a constant slope near that value which is also inconsistent with the slope derived from the PR relation.

Using the F--test, there is evidence for non--linearity in the relations similar to the non--linear structure found in optical PL relations.  Mass loss acts to linearize the PL relation at $3.6,4.5,$ and $5.8$ $\mu m$, while at $8.0$ $\mu m$ the PL relation is non--linear with a slope that is shallower at $P<10d$ than for $P>10d$ which implies the infrared excess is becoming more important at longer wavelength.  Mass loss may also explain why the K--band PL relation is marginally linear \cite{Ngeow2006}. 

The resulting effect that mass loss has on infrared observations of Cepheids implies serious consequences for infrared Period--Luminosity relations if they are to be used for high precision astrophysics.  One of the reasons for using infrared PL relations is that they are less sensitive to metallicity than optical relations and hence do not need to be corrected for each galaxy \citep{Sasselov1997}.  The metallicity correction is a significant source of uncertainty in studies of the Hubble Constant \citep{Freedman2001} and an infrared PL relation that avoids this uncertainty would be a powerful tool.  However, we have shown that mass loss affects the scatter and the structure of the PL relation.  The scatter increases the uncertainty of any distance determination, but more importantly the fractional amount of dust generated in a wind depends on metallicity.  The amount of mass loss may also depend on metallicity, as suggested by \cite{Neilson2008b}.  These two issues imply the Period--Luminosity relation depends on metallicity at infrared wavelengths as well as at optical wavelengths though to what extent is currently unknown.    

\begin{acknowledgements}
HRN is grateful for funding from the Walter John Helm OGSST and the Walter C. Sumner Memorial Fellowship.
\end{acknowledgements} 
\bibliography{wind_th}

\providecommand{\noopsort}[1]{}
\begin{thebibliography}{38}
\expandafter\ifx\csname natexlab\endcsname\relax\def\natexlab#1{#1}\fi

\bibitem[{{Beaulieu} {et~al.}(2001){Beaulieu}, {Buchler}, \&
  {Koll{\'a}th}}]{Beaulieu2001}
{Beaulieu}, J.~P., {Buchler}, J.~R., \& {Koll{\'a}th}, Z. 2001, \aap, 373, 164

\bibitem[{{Bono} {et~al.}(2000){Bono}, {Castellani}, \& {Marconi}}]{Bono2000}
{Bono}, G., {Castellani}, V., \& {Marconi}, M. 2000, \apj, 529, 293

\bibitem[{{Brocato} {et~al.}(2004){Brocato}, {Caputo}, {Castellani}, {Marconi},
  \& {Musella}}]{Brocato2004}
{Brocato}, E., {Caputo}, F., {Castellani}, V., {Marconi}, M., \& {Musella}, I.
  2004, \aj, 128, 1597

\bibitem[{{Caputo} {et~al.}(2005){Caputo}, {Bono}, {Fiorentino}, {Marconi}, \&
  {Musella}}]{Caputo2005}
{Caputo}, F., {Bono}, G., {Fiorentino}, G., {Marconi}, M., \& {Musella}, I.
  2005, \apj, 629, 1021

\bibitem[{{Castor} {et~al.}(1975){Castor}, {Abbott}, \& {Klein}}]{Castor1975}
{Castor}, J.~I., {Abbott}, D.~C., \& {Klein}, R.~I. 1975, \apj, 195, 157

\bibitem[{{Catelan} \& {Cort{\'e}s}(2008)}]{Catelan2008}
{Catelan}, M. \& {Cort{\'e}s}, C. 2008, \apjl, 676, L135

\bibitem[{{Clayton} \& {Martin}(1985)}]{Clayton1985}
{Clayton}, G.~C. \& {Martin}, P.~G. 1985, \apj, 288, 558

\bibitem[{{Clement} {et~al.}(2008){Clement}, {Xu}, \& {Muzzin}}]{Clement2008}
{Clement}, C.~M., {Xu}, X., \& {Muzzin}, A.~V. 2008, \aj, 135, 83

\bibitem[{{Deasy}(1988)}]{Deasy1988}
{Deasy}, H.~P. 1988, \mnras, 231, 673

\bibitem[{{Freedman} {et~al.}(2001){Freedman}, {Madore}, {Gibson}, {Ferrarese},
  {Kelson}, {Sakai}, {Mould}, {Kennicutt}, {Ford}, {Graham}, {Huchra},
  {Hughes}, {Illingworth}, {Macri}, \& {Stetson}}]{Freedman2001}
{Freedman}, W.~L., {Madore}, B.~F., {Gibson}, B.~K., {Ferrarese}, L., {Kelson},
  D.~D., {Sakai}, S., {Mould}, J.~R., {Kennicutt}, Jr., R.~C., {Ford}, H.~C.,
  {Graham}, J.~A., {Huchra}, J.~P., {Hughes}, S.~M.~G., {Illingworth}, G.~D.,
  {Macri}, L.~M., \& {Stetson}, P.~B. 2001, \apj, 553, 47

\bibitem[{{Freedman} {et~al.}(2008){Freedman}, {Madore}, {Rigby}, {Persson}, \&
  {Sturch}}]{Freedman2008}
{Freedman}, W.~L., {Madore}, B.~F., {Rigby}, J., {Persson}, S.~E., \& {Sturch},
  L. 2008, \apj, 679, 71

\bibitem[{{Gieren} {et~al.}(1999){Gieren}, {Moffett}, \& {Barnes}}]{Gieren1999}
{Gieren}, W.~P., {Moffett}, T.~J., \& {Barnes}, III, T.~G. 1999, \apj, 512, 553

\bibitem[{{Kanbur} \& {Ngeow}(2004{\natexlab{a}})}]{Kanbur2004}
{Kanbur}, S.~M. \& {Ngeow}, C.-C. 2004{\natexlab{a}}, \mnras, 350, 962

\bibitem[{{Kanbur} \& {Ngeow}(2004{\natexlab{b}})}]{Kanbur2004a}
---. 2004{\natexlab{b}}, \mnras, 350, 962

\bibitem[{{Kanbur} {et~al.}(2004){Kanbur}, {Ngeow}, \& {Buchler}}]{Kanbur2004b}
{Kanbur}, S.~M., {Ngeow}, C.-C., \& {Buchler}, J.~R. 2004, \mnras, 354, 212

\bibitem[{{Keller}(2008)}]{Keller2008}
{Keller}, S.~C. 2008, \apj, 677, 483

\bibitem[{{Keller} \& {Wood}(2006)}]{Keller2006}
{Keller}, S.~C. \& {Wood}, P.~R. 2006, \apj, 642, 834

\bibitem[{{Kervella} {et~al.}(2006){Kervella}, {M{\'e}rand}, {Perrin}, \&
  {Coud{\'e} Du Foresto}}]{Kervella2006}
{Kervella}, P., {M{\'e}rand}, A., {Perrin}, G., \& {Coud{\'e} Du Foresto}, V.
  2006, \aap, 448, 623

\bibitem[{{Kurucz}(1979)}]{Kurucz1979}
{Kurucz}, R.~L. 1979, \apjs, 40, 1

\bibitem[{{Lah} {et~al.}(2005){Lah}, {Kiss}, \& {Bedding}}]{Lah2005}
{Lah}, P., {Kiss}, L.~L., \& {Bedding}, T.~R. 2005, \mnras, 359, L42

\bibitem[{{Lamers} \& {Cassinelli}(1999)}]{Lamers1999}
{Lamers}, H.~J.~G.~L.~M. \& {Cassinelli}, J.~P. 1999, {Introduction to Stellar
  Winds} (Introduction to Stellar Winds, by Henny J.~G.~L.~M.~Lamers and Joseph
  P.~Cassinelli, pp.~452.~ISBN 0521593980.~Cambridge, UK: Cambridge University
  Press, June 1999.)

\bibitem[{{Laney} \& {Stobie}(1994)}]{Laney1994}
{Laney}, C.~D. \& {Stobie}, R.~S. 1994, \mnras, 266, 441

\bibitem[{{Mathis} {et~al.}(1977){Mathis}, {Rumpl}, \&
  {Nordsieck}}]{Mathis1977}
{Mathis}, J.~S., {Rumpl}, W., \& {Nordsieck}, K.~H. 1977, \apj, 217, 425

\bibitem[{{Mattsson} {et~al.}(2008){Mattsson}, {Wahlin}, {H{\"o}fner}, \&
  {Eriksson}}]{Mattsson2008}
{Mattsson}, L., {Wahlin}, R., {H{\"o}fner}, S., \& {Eriksson}, K. 2008, ArXiv
  e-prints, 804

\bibitem[{{McAlary} \& {Welch}(1986)}]{McAlary1986}
{McAlary}, C.~W. \& {Welch}, D.~L. 1986, \aj, 91, 1209

\bibitem[{{Meixner} {et~al.}(2006){Meixner}, {Gordon}, {Indebetouw}, {Hora},
  {Whitney}, {Blum}, {Reach}, {Bernard}, {Meade}, {Babler}, {Engelbracht},
  {For}, {Misselt}, {Vijh}, {Leitherer}, {Cohen}, {Churchwell}, {Boulanger},
  {Frogel}, {Fukui}, {Gallagher}, {Gorjian}, {Harris}, {Kelly}, {Kawamura},
  {Kim}, {Latter}, {Madden}, {Markwick-Kemper}, {Mizuno}, {Mizuno}, {Mould},
  {Nota}, {Oey}, {Olsen}, {Onishi}, {Paladini}, {Panagia}, {Perez-Gonzalez},
  {Shibai}, {Sato}, {Smith}, {Staveley-Smith}, {Tielens}, {Ueta}, {Dyk},
  {Volk}, {Werner}, \& {Zaritsky}}]{Meixner2006}
{Meixner}, M., {Gordon}, K.~D., {Indebetouw}, R., {Hora}, J.~L., {Whitney}, B.,
  {Blum}, R., {Reach}, W., {Bernard}, J.-P., {Meade}, M., {Babler}, B.,
  {Engelbracht}, C.~W., {For}, B.-Q., {Misselt}, K., {Vijh}, U., {Leitherer},
  C., {Cohen}, M., {Churchwell}, E.~B., {Boulanger}, F., {Frogel}, J.~A.,
  {Fukui}, Y., {Gallagher}, J., {Gorjian}, V., {Harris}, J., {Kelly}, D.,
  {Kawamura}, A., {Kim}, S., {Latter}, W.~B., {Madden}, S., {Markwick-Kemper},
  C., {Mizuno}, A., {Mizuno}, N., {Mould}, J., {Nota}, A., {Oey}, M.~S.,
  {Olsen}, K., {Onishi}, T., {Paladini}, R., {Panagia}, N., {Perez-Gonzalez},
  P., {Shibai}, H., {Sato}, S., {Smith}, L., {Staveley-Smith}, L., {Tielens},
  A.~G.~G.~M., {Ueta}, T., {Dyk}, S.~V., {Volk}, K., {Werner}, M., \&
  {Zaritsky}, D. 2006, \aj, 132, 2268

\bibitem[{{M{\'e}rand} {et~al.}(2007){M{\'e}rand}, {Aufdenberg}, {Kervella},
  {Foresto}, {ten Brummelaar}, {McAlister}, {Sturmann}, {Sturmann}, \&
  {Turner}}]{Merand2007}
{M{\'e}rand}, A., {Aufdenberg}, J.~P., {Kervella}, P., {Foresto}, V.~C.~d.,
  {ten Brummelaar}, T.~A., {McAlister}, H.~A., {Sturmann}, L., {Sturmann}, J.,
  \& {Turner}, N.~H. 2007, \apj, 664, 1093

\bibitem[{{M{\'e}rand} {et~al.}(2006){M{\'e}rand}, {Kervella}, {Coud{\'e} Du
  Foresto}, {Perrin}, {Ridgway}, {Aufdenberg}, {Ten Brummelaar}, {McAlister},
  {Sturmann}, {Sturmann}, {Turner}, \& {Berger}}]{Merand2006}
{M{\'e}rand}, A., {Kervella}, P., {Coud{\'e} Du Foresto}, V., {Perrin}, G.,
  {Ridgway}, S.~T., {Aufdenberg}, J.~P., {Ten Brummelaar}, T.~A., {McAlister},
  H.~A., {Sturmann}, L., {Sturmann}, J., {Turner}, N.~H., \& {Berger}, D.~H.
  2006, \aap, 453, 155

\bibitem[{{Neilson} \& {Lester}(2008{\natexlab{a}})}]{Neilson2008}
{Neilson}, H.~R. \& {Lester}, J.~B. 2008{\natexlab{a}}, \apj, 684, 569

\bibitem[{{Neilson} \& {Lester}(2008{\natexlab{b}})}]{Neilson2008b}
---. 2008{\natexlab{b}}, ArXiv e-prints

\bibitem[{{Ngeow} \& {Kanbur}(2006)}]{Ngeow2006}
{Ngeow}, C. \& {Kanbur}, S.~M. 2006, \apj, 650, 180

\bibitem[{{Ngeow} \& {Kanbur}(2008)}]{Ngeow2008}
---. 2008, \apj, 679, 76

\bibitem[{{Ngeow} {et~al.}(2005){Ngeow}, {Kanbur}, {Nikolaev}, {Buonaccorsi},
  {Cook}, \& {Welch}}]{Ngeow2005}
{Ngeow}, C.-C., {Kanbur}, S.~M., {Nikolaev}, S., {Buonaccorsi}, J., {Cook},
  K.~H., \& {Welch}, D.~L. 2005, \mnras, 363, 831

\bibitem[{{Sasselov} {et~al.}(1997){Sasselov}, {Beaulieu}, {Renault}, {Grison},
  {Ferlet}, {Vidal-Madjar}, {Maurice}, {Prevot}, {Aubourg}, {Bareyre},
  {Brehin}, {Coutures}, {Delabrouille}, {de Kat}, {Gros}, {Laurent},
  {Lachieze-Rey}, {Lesquoy}, {Magneville}, {Milsztajn}, {Moscoso}, {Queinnec},
  {Rich}, {Spiro}, {Vigroux}, {Zylberajch}, {Ansari}, {Cavalier}, {Moniez},
  {Gry}, {Guibert}, {Moreau}, \& {Tajhmady}}]{Sasselov1997}
{Sasselov}, D.~D., {Beaulieu}, J.~P., {Renault}, C., {Grison}, P., {Ferlet},
  R., {Vidal-Madjar}, A., {Maurice}, E., {Prevot}, L., {Aubourg}, E.,
  {Bareyre}, P., {Brehin}, S., {Coutures}, C., {Delabrouille}, N., {de Kat},
  J., {Gros}, M., {Laurent}, B., {Lachieze-Rey}, M., {Lesquoy}, E.,
  {Magneville}, C., {Milsztajn}, A., {Moscoso}, L., {Queinnec}, F., {Rich}, J.,
  {Spiro}, M., {Vigroux}, L., {Zylberajch}, S., {Ansari}, R., {Cavalier}, F.,
  {Moniez}, M., {Gry}, C., {Guibert}, J., {Moreau}, O., \& {Tajhmady}, F. 1997,
  \aap, 324, 471

\bibitem[{{Udalski} {et~al.}(1999b){Udalski}, {Soszynski}, {Szymanski},
  {Kubiak}, {Pietrzynski}, {Wozniak}, \& {Zebrun}}]{Udalski1999b}
{Udalski}, A., {Soszynski}, I., {Szymanski}, M., {Kubiak}, M., {Pietrzynski},
  G., {Wozniak}, P., \& {Zebrun}, K. 1999b, Acta Astronomica, 49, 223

\bibitem[{{Udalski} {et~al.}(1999a){Udalski}, {Szymanski}, {Kubiak},
  {Pietrzynski}, {Soszynski}, {Wozniak}, \& {Zebrun}}]{Udalski1999a}
{Udalski}, A., {Szymanski}, M., {Kubiak}, M., {Pietrzynski}, G., {Soszynski},
  I., {Wozniak}, P., \& {Zebrun}, K. 1999a, Acta Astronomica, 49, 201

\bibitem[{{Weingartner} \& {Draine}(2001)}]{Weingartner2001}
{Weingartner}, J.~C. \& {Draine}, B.~T. 2001, \apj, 548, 296

\bibitem[{{Willson}(1989)}]{Willson1989}
{Willson}, L.~A. 1989, in IAU Colloq. 111: The Use of pulsating stars in
  fundamental problems of astronomy, ed. E.~G. {Schmidt}, 63--+

\end{thebibliography}
\bibliographystyle{apj}

\end{document}